\documentclass[aps,preprint]{revtex4}

\usepackage{graphics,epsfig,amssymb}

\def\beq{\begin{equation}} 
\def\eeq{\end{equation}} 
\begin{document}

\title{Quartet condensation induced by the isovector pairing force}

\author{M. Sambataro$^a$ and N. Sandulescu$^b$}
\affiliation{$^a$Istituto Nazionale di Fisica Nucleare - Sezione di Catania,
Via S. Sofia 64, I-95123 Catania, Italy \\
$^b$National Institute of Physics and Nuclear Engineering, P.O. Box MG-6, 
Magurele, Bucharest, Romania}

\begin{abstract}
The phenomenon of quartet condensation in the ground state of an isovector pairing Hamiltonian for an even-even $N=Z$ system is investigated. For this purpose we follow the 
evolution of the ground state from an unperturbed regime up to a strongly interacting one in a formalism of collective pairs. These pairs are those resulting from the diagonalization of the pairing Hamiltonian in a space of two particles coupled to isospin $T=1$. The ground state is found to rapidly evolve from a product of distinct $T=0$ quartets, each one formed by two of the above pairs, to a condensate of identical quartets built only with the pair corresponding to the lowest energy. This finding establishes a link between the complicated structure of  the exact ground state and the simple approximation scheme of the Quartet Condensation Model. The mechanism at the basis of this quartet condensation turns out to be the same which is
responsible for the development of a pair condensate in the ground state of a like-particle pairing Hamiltonian.
\end{abstract}

\maketitle

\section{Introduction}
In a recent paper \cite{sasa_jpg}, we have proposed a new derivation of the exact $T=0$ seniority-zero eigenstates of the isovector pairing Hamiltonian for an even-even $N=Z$ system. The exact solution of this Hamiltonian, which plays a key role in the comprehension of proton-neutron correlations in nuclear systems \cite{frau}, has represented a longstanding issue in nuclear structure. A complete exact treatment of this Hamiltonian (including also isospin symmetry-breaking terms) was presented in 2006 by Dukelsky et al. \cite{duke} but previous attempts date back up to the 60's with the pioneering work of Richardson \cite{richa1} which turned out, however, to be only partially correct \cite{feng}.

The basic goal of the approach of Ref. \cite{sasa_jpg} has been that of providing a physically transparent description of the exact eigenstates of the isovector pairing Hamiltonian. It has been shown that these eigenstates can be formulated as linear superpositions of products of isospin $T=0$ quartets, each quartet being formed by two collective $T=1$ pairs. The structure of these collective pairs has turned out to be  identical to that of the pairs introduced by Richardson in the exact treatment of the like-particle pairing Hamiltonian \cite{richa0,richa3,richa2}. 

In spite of their conceptually simple structure, the eigenstates which have been constructed in 
Ref. \cite{sasa_jpg} are nevertheless quite complicated. This is due to the large number of components in the expansion (e.g., 105 for a system of 8 protons and 8 neutrons distributed over 8 levels) and to
the fact that both the amplitudes of the pairs and the coefficients of this expansion can be either real or complex. Furthermore, no simplification in the structure of the ground state has been observed with varying the pairing strength. The only exception is represented by the extreme case of a pairing strength close 
to zero, where the ground state reduces itself to a single product of distinct quartets.

In previous works \cite{sandu,sasa_prc}, it has been shown that the ground state of an isovector 
pairing Hamiltonian for an even-even $N=Z$ system can be approximated to a very high degree of accuracy as a single product of identical $T=0$ quartets. These quartets are built by two collective isovector pairs 
coupled to the total isospin $T=0$. This approach, known as the Quartet Condensation Model (QCM), has 
been extended later on to the treatment of isoscalar-isovector  pairing interactions  \cite{qcm_t0t1,qm_t0t1,sasa_t01} and also to more general two-body interactions of shell model type
\cite{qcm_epj}. The QCM wave function is by far simpler than the exact one and its link with the latter is anything but obvious. Understanding this link will shed light on the phenomenon of quartet condensation in the ground state of the isovector pairing Hamiltonian. This will be the goal of
the present work.

The relation between the exact solution and the quartet condensation discussed in this paper
will be contrasted with an analogous relation for a state-independent pairing interaction
between like particles. In 1963 Richardson showed that in the latter case the exact 
ground state is a product of distinct collective pairs (either real or complex)\cite{richa0,richa3,richa2}.
A few years earlier Bayman \cite{bayman} had proposed, for the same Hamiltonian, the so-called
PBCS approximation in which the ground state is supposed to be a pair condensate, i.e.,
a product of identical collective pairs (only real). This approximation has turned out to be quite accurate 
in spite of its simplicity \cite{sandu2,dukel}. The phenomenon of pair condensation in the ground state of the like-particle pairing has been investigated in Ref. \cite{samba}.
In the present work we will show that this phenomenon and that of 
quartet condensation in the case of the isovector pairing share a common origin.

\section{Pair condensation and the like-particle pairing}
To highlight the analogy between the emergency of pair and quartet condensates, 
in this section we will recall and further elaborate a few
results from Ref. \cite{samba} which concern the case of like-particle pairing. 

We consider the Hamiltonian
\begin{equation}
H=\sum^{\Omega}_{j=1}\epsilon_jN_j-g\sum^{\Omega}_{i,j=1}P^{\dag}_iP_j,
\label{1}
\end{equation}
where
\begin{equation}
N_j=\sum_\sigma a^{\dag}_{j\sigma}a_{j\sigma},~~~
P^{\dag}_j=a^{\dag}_{j+}a^{\dag}_{j-},~~~ P_j=(P^{\dag}_j)^{\dag}.
\end{equation}
This Hamiltonian describes a system of fermions
distributed over a set of $\Omega$ doubly degenerate single particle levels and 
interacting via a pairing force with a level-independent strength $g$.
The operator $a^{\dag}_{j\sigma}$ ($a_{j\sigma}$) creates (annihilates)
a fermion in the 
single-particle state $(j,\sigma )$, where $j$ identifies one of the $\Omega$
levels of the model and $\sigma =\pm$ labels time reversed states.

The derivation of the exact eigenstates and eigenvalues of the
Hamiltonian (\ref{1}) has been extensively discussed in Refs.
\cite{richa0,richa3,richa2}. Here, we only recall that the $N$-pair state
\begin{equation}
|\Psi\rangle =\prod^{N}_{i=1}B^{\dag}_i|0\rangle,~~~~~
B^{\dag}_i=\sum^{\Omega}_{k=1}\frac{1}{2\epsilon_k-\lambda_i}P^{\dag}_k
\label{2}
\end{equation}
is a seniority-zero
eigenstate of the Hamiltonian (\ref{1}) if the $N$ parameters $\lambda_i$ associated with this state are roots of
the set of $N$ coupled non-linear equations
\begin{equation}
1-\sum^{\Omega}_{k=1}\frac{g}{2\epsilon_k-\lambda_i}+
\sum^{N}_{l(l\neq i)=1}\frac{2g}{\lambda_l-\lambda_i}=0.
\label{3}
\end{equation}
The eigenvalue $E^{(\Psi )}$ associated with $|\Psi\rangle$ turns out to be the sum
of the $\lambda_i$'s, 
\begin{equation}
E^{(\Psi )}=\sum^{N}_{i=1}\lambda_i. 
\label{4}
\end{equation}
The parameters  $\lambda_i$ exhibit a peculiar behavior. With increasing the strength $g$, they become equal two by two  and, when this occurs, they turn from real into complex-conjugate pairs. The equality of two $\lambda_i$'s gives rise to a singularity in Eqs. (\ref{3}) which can be handled with an appropriate change of variables. 
The qualitative behavior of these parameters in the ground state is illustrated in Ref. \cite{richa2}. 

The parameters $\lambda_i$ by themselves have no physical meaning, except for the case of a one-pair system.
For this system $\lambda_i$ coincides with the energy of the pair $B^\dag_i$ . This can be seen from the expression 
\begin{equation}
HB^\dag_i|0\rangle =\lambda_i B^\dag_i|0\rangle 
+(1-g\sum_k\frac{1}{2\epsilon_k-\lambda_i})(\sum_jP^{\dag}_j)|0\rangle ,
\label{5}
\end{equation}
which shows that  $B^\dag_i|0\rangle$ is an eigenstate of $H$ with the eigenvalue $\lambda_i$ 
if this quantity satisfies the condition
\begin{equation}
1-g\sum_k\frac{1}{2\epsilon_k-\lambda_i}=0 .
\label{6}
\end{equation}
This condition is nothing but the particular form of Richardson equations (\ref{3}) for a one-particle system. 

It should be noticed that for a general $N$-pair system the $\lambda_i$'s satisfy the conditions (\ref{3}) so that the state $B^\dag_i|0\rangle$, formed with one of the $N$ Richardson pairs (\ref{2}), is not an eigenstate of the Hamiltonian (1) and the related $\lambda_i$ is not the corresponding energy.
In the following, to each Richardson pair $B^\dag_i$ we will associate the energy $E_i=\langle 0|B_iHB^\dag_i|0\rangle/\langle 0|B_iB^\dag_i|0\rangle$.
This quantity represents the energy of two particles in the state $B^\dag_i| 0 \rangle$ which interact through the Hamiltonian (1).

\begin{figure}
\includegraphics{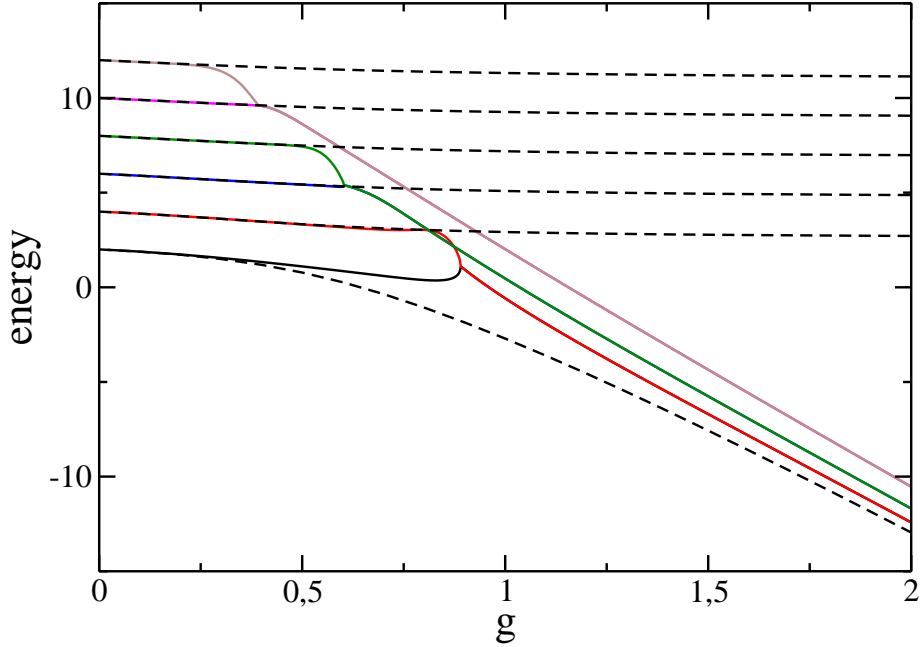}
\mbox{}\\[9.0cm]
\caption{Solid lines show the energies of the pairs $B^\dag_i$ (\ref{2}) defining the ground state of $H$ (\ref{1}) for a system with $\Omega =2N=12$.
Dashed lines show the energies of the lowest six pairs $\Gamma^\dag_i$ resulting from the diagonalization of $H$ in the space spanned by the states $P^\dag_k|0\rangle$ (see text).
All values are in units of the level spacing $d$.}
\label{fig1}
\end{figure}

To illustrate the properties of the energies $E_i$ we  examine the case of a (half-filled) system with 12 particles distributed over 12 equispaced levels (i.e. $\epsilon_j=jd$, $d$ being the level spacing).
In Fig. \ref{fig1}, the solid lines show the energies $E_i$ of the pairs $B^\dag_i$ (\ref{2}) which define the ground state of this system. The behavior of these energies as a function of the strength $g$ closely reminds that of the corresponding parameters $\lambda_i$ \cite{samba}. 
They approach the values $2\epsilon_i$ for $g\rightarrow 0$ while, with increasing $g$, the $E_i$'s become equal two by two, starting from the highest ones in energy. This occurs at the same critical points where the  $\lambda_i$'s become equal. After each critical point, the energies $E_i$ corresponding to pairs with complex-conjugate $\lambda_i$ become equal (by remaining always real) and
then they decrease with almost constant slope. Interestingly, after the last critical point these energies stay close to each other, indicating that the corresponding pairs have similar correlations properties. 

In Fig. \ref{fig1} are also shown, by  dashed lines, the energies of the lowest 6 pairs 
which result from the diagonalization of $H$ in the $N=1$ space spanned by the states $P^\dag_k|0\rangle$. These pairs, which satisfy Eqs. (\ref{5}) and (\ref{6}), will be denoted in the following by $\Gamma^\dag_i$  to distinguish them from the Richardson pairs $B^\dag_i$ (\ref{2}) defining the eigenstates of a generic $N$-pair system.
One observes a marked difference between the behavior of the
lowest energy and that of the remaining ones, with the former decreasing quite rapidly with incresing $g$ and all others (included the six uppermost energies not shown in the figure) staying almost constant. This different behavior reflects a marked difference in the structure of the pairs $\Gamma^\dag_i$, the lowest pair being the only one whose amplitudes carry all the same sign and so being also the pair with the maximum degree of coherence among its components \cite{samba}.

There are two features of Fig. \ref{fig1} which deserve to be noticed. The first feature concerns the joint behaviour of the dashed and solid lines before the critical points. Within given intervals of $g$, different for each pair but never exceeding the critical point associated with that pair, these lines perfectly overlap each other. This denotes a close identity, within these intervals, between the Richardson pairs $B^\dag_i$ defining the ground state (\ref{2}) and the lowest pairs $\Gamma^\dag_i$ resulting from the diagonalization of $H$ in the space $\{P^\dag_k|0\rangle \}$. For $g$ smaller than the first critical point, approximately $0\leq g\leq 0.25$, all the 6 pairs $B^\dag_i$ in the exact ground state (\ref{3}) essentially coincide with the 6 lowest pairs $\Gamma^\dag_i$. The second very interesting feature seen in Fig. \ref{fig1} is the convergence of the lowest dashed line to the energies
of the Richardson pairs in the strong coupling regime. This fact indicates that, in this regime, the Richardson pairs have correlations properties similar to those of the lowest collective pair of the one-pair system.

\begin{figure}
\includegraphics{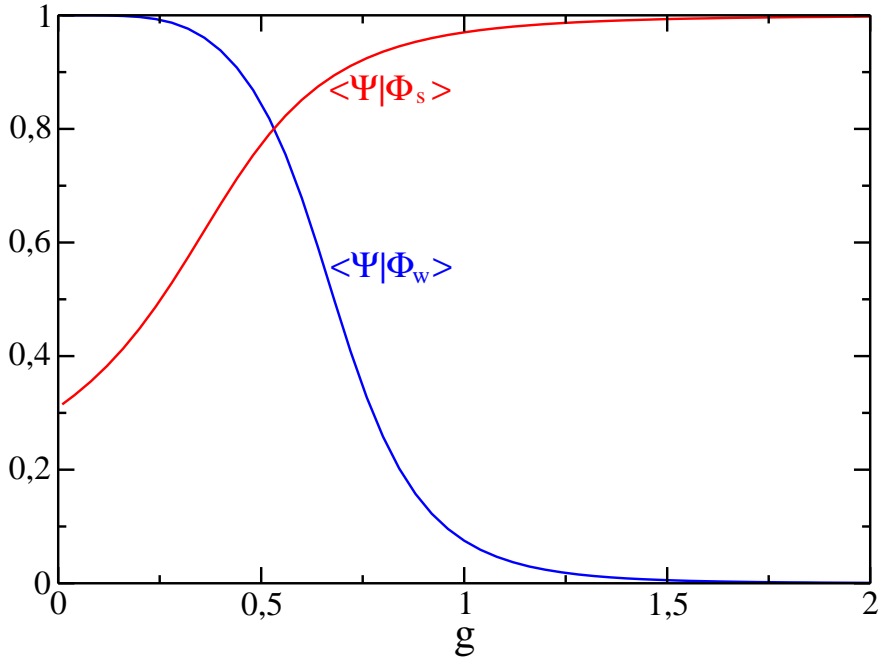}
\mbox{}\\[9.0cm]
\caption{The overlaps $\langle\Psi |\Phi_w\rangle$ and $\langle\Psi |\Phi_s\rangle$
discussed in the text as a function of the strength $g$ (in units of the level spacing $d$). Data are from Ref. \cite{samba}.}
\label{fig2}
\end{figure}

The 12 collective pairs $\Gamma^\dag_i$ can be used to construct a set of basis states $\{\Gamma^\dag_{i_1}\Gamma^\dag_{i_2}\cdot\cdot\cdot\Gamma^\dag_{i_6}|0\rangle \}$ (where $1\leq i_1\leq i_2\cdot\cdot\cdot\leq i_6\leq 12$) in terms of which to expand the exact eigenstates of $H$. By ordering the pairs $\Gamma^\dag_i$ for increasing values of their energies and following the previous discussion, one expects an overlap close to unity between the exact ground state $|\Psi\rangle$ and the state 
$|\Phi_w\rangle =\Gamma^\dag_1\Gamma^\dag_2\cdot\cdot\cdot\Gamma^\dag_6|0\rangle$
(properly normalized) in the weak coupling region $0\leq g\leq 0.25$. This can indeed be seen in Fig. \ref{fig2}. With increasing $g$, this overlap rapidly decreases to zero for $g\geq 1.25$. In the same figure one finds the overlap between the exact ground state $|\Psi\rangle$ and the state $|\Phi_s\rangle =(\Gamma^\dag_1)^6|0\rangle$, namely a condensate of the lowest pair $\Gamma^\dag_1$. This overlap behaves oppositely to the previous one, being pretty small in the weak coupling region and approaching one in the strong coupling region $g\geq 1.25$. Thus, with increasing $g$, in spite of its complex structure (\ref{2}), the ground state approaches the form of a pair condensate built with the lowest pair $\Gamma^\dag_1$. The reason why this pair condensate becomes predominant over all
other states that can be built with the pairs $\Gamma^\dag_i$ can be identified in the rapid increase with the pairing strength of the energy gap between the lowest pair  and the remaining ones.

We remark that the effectiveness of the condensate  $|\Phi_s\rangle$ can be improved considerably by replacing the pair  $\Gamma^\dag_1$ with a generic collective pair 
determined from the minimisation of  $\langle \Phi_s |H| \Phi_s \rangle$.
This is what is commonly done in the standard PBCS approximation. How the collectivity of the PBCS and Richardson pairs are related to each other is discussed in Ref. \cite{sandu2}

The properties of the Richardson pairs ($\ref{2}$) and the associated energies in the strong coupling regime
point out an even better approximation than the pair condensation discussed above. According to
the exact solution, with increasing $g$, the Richardson pairs become two-by-two complex conjugates,
i.e., they correspond to complex conjugates parameters $\lambda_k=\xi_k \pm i\eta_k$. It can be easily
shown that the product of two complex conjugates pairs can be written as a real ``quartet", i.e., a 4-body 
correlated structure characterized by real amplitudes \cite{qcm_su2}. Therefore, in the strong coupling regime, when all the 
pairs have become complex conjugates, the exact ground state can be represented by a product of distinct real
quartets. On the other hand, as seen in Fig.\ref{fig1}, in the strong coupling regime the energies
of the complex conjugates pairs which compose the quartets are close to each other, indicating that
the quartets have similar properties. This fact suggests that the ground state can
be approximated by a condensate of quartets, an approach which we have introduced and studied 
in Ref. \cite{qcm_su2}. It has been shown that the quartet condensation is a better approximation than 
the pair condensation. However, the extra correlation energy gained by the quartet
condensation is not significantly large.

\section{Quartet condensation and the isovector pairing}
The Hamiltonian that we examine in this section has the form
\begin{equation}
H^{(iv)}=\sum^\Omega_{i=1}\epsilon_i{\cal N}_i-g\sum^\Omega_{i,i'=1}\sum^{1}_{M_T=-1}P^{\dag}_{iM_T}P_{i'M_T},
\label{7}
\end{equation}
where
\begin{equation}
{\cal N}_i=\sum_{\sigma =\pm ,\tau =\pm\frac{1}{2}}a^\dag_{i\sigma\tau}a_{i\sigma \tau},~~~~
P^\dag_{iM_T}=[a^\dag_{i+}a^\dag_{i-}]^{T=1}_{M_T}.
\label{8}
\end{equation}
This Hamiltonian describes a system of protons and neutrons distributed over a set of $\Omega$ levels and interacting via an isovector pairing force with a level-independent strength $g$.
The operator $a^\dag_{i\sigma\tau}$ 
($a_{i\sigma\tau}$) creates (annihilates) a nucleon in the single-particle state characterized by the quantum numbers
$(i, \sigma ,\tau )$, where $i$ identifies one of the $\Omega $ levels of the model, 
$\sigma =\pm$ labels states which are conjugate with respect to time reversal and
$\tau =\pm\frac{1}{2}$ is the projection of the isospin of the nucleon.
The operator $P^\dag_{iM_T}$ $(P_{iM_T})$ creates (annihilates) a pair of nucleons in time-reversed states with total isospin $T=1$ and projection $M_T$. Depending on $M_T$, $P^\dag_{iM_T}$ creates a $pp$, a $nn$ or a $pn$ pair and the Hamiltonian $(\ref{1})$ is seen to act equally on these pairs.
Finally the operator ${\cal N}_i$ counts the number of nucleons on the level $i$, each level having an energy $\epsilon_i$. 

Owing to the presence of the $\sigma ,  \tau$ degrees of freedom, each level $i$ is fourfold degenerate being able to accommodate two protons and two neutrons in time-reversed states. We limit the Hilbert space of the model to seniority-zero states  \cite{richa1}. 
For a $2N$-particle system this space is spanned by the states
\begin{equation}
P^\dag_{i_1M_{T_1}}P^\dag_{i_2M_{T_2}}\cdots P^\dag_{i_NM_{T_N}}|0\rangle ,
\label{9}
\end{equation}
where $|0\rangle$ is the vacuum of the model. Having in mind a description of $T=0$ states only, we further require that 
$M_{T_1}+M_{T_2}+\dots +M_{T_N}=0$. 

It has been shown in Ref. \cite{sasa_jpg} that any seniority-zero $T=0$ eigenstate 
of $H^{(iv)}$ can be expressed as a linear superposition of products of $T=0$ quartets
formed by two collective $T=1$ pairs, namely
\begin{equation}
|\Psi^{(iv)}\rangle =\sum^{N_s}_{s=1}d_s|s\rangle ,~~~~~
|s\rangle =\prod^{N/2}_{q=1}[B^\dagger_{\nu (1,q,s)}B^\dagger_{\nu (2,q,s)}]^{T=0}|0\rangle ,
\label{10}
\end{equation}
with
\begin{equation}
B^{\dag}_{\nu M_T}=\sum^{\Omega}_{k=1}\frac{1}{2\epsilon_k-\lambda_{\nu}}P^{\dag}_{kM_T} .
\label{11}
\end{equation}
The collective pairs $B^{\dag}_{\nu M_T}$  are characterized by the same amplitudes of the pairs (\ref{2}). As in Eq. (\ref{4}),
the eigenvalue corresponding to $|\Psi^{(iv)}\rangle$ is the sum of the parameters $\lambda_{\nu}$  associated with this state.

The eigenstate (\ref{10}) depends on two sets of variables, the coefficients $d_s$ of the expansion and the parameters $\lambda_\nu$, which are found by solving the system of coupled non-linear equations
\begin{equation}
\frac{d_s}{g}-\sum_i\frac{d_s}{2\epsilon_i-\lambda_{\nu}}-
\sum_{\nu '\neq\nu}\frac{S_{\nu '\nu}(s)}{\lambda_{\nu '}-\lambda_\nu}=0.
\label{12}
\end{equation}
Details on the matrices $S_{\nu '\nu}(s)$ and on the solution of these equations can be found in Ref. \cite{sasa_jpg}. Here we only remind that 
both the $\lambda_\nu$'s and the coefficients $d_s$ of the expansion (\ref{10}) can be either real or complex. As for the like-particle pairing, at some critical values of the pairing strength, two (real) 
$\lambda_{\nu}$'s become equal thus giving rise to a singularity. At this point, they  turn from real  into complex-conjugates pairs.

We shall consider a system with 12 particles, 6 protons and 6 neutrons, at half-filling
as in the previous section. Because of the fourfold degeneracy of the levels, this implies distributing the particles over 6 levels. We shall adopt the same single-particle energies as in Ref. \cite{sasa_jpg}, namely $\epsilon_i=-16+2(i-1)$, which are characterized by a constant spacing $d=2$. 
\begin{figure}
\includegraphics{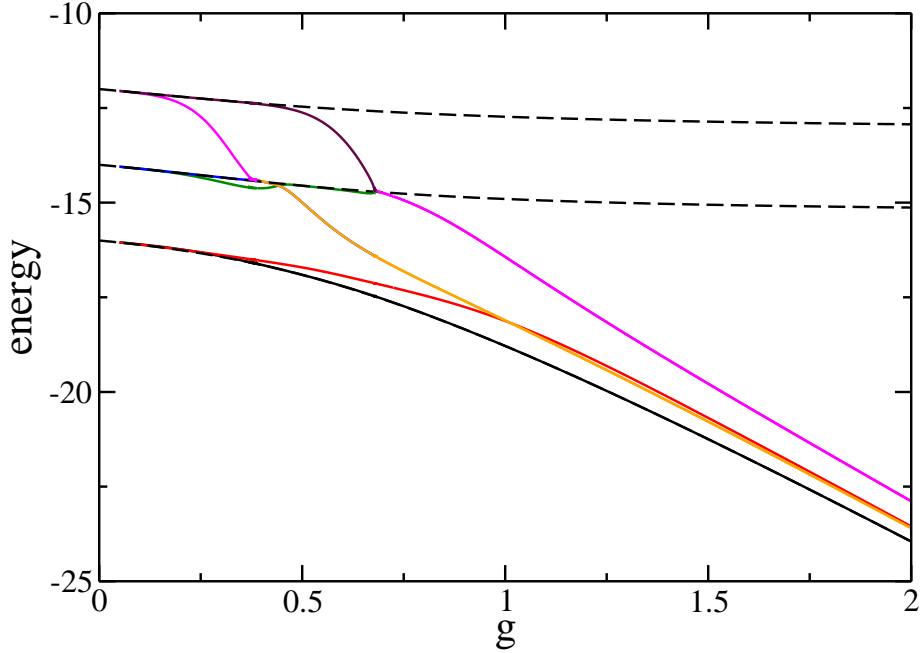}
\mbox{}\\[9.0cm]
\caption{Solid lines show the energies of the pairs $B^\dag_\nu$ (\ref{11}) defining the ground state of $H^{(iv)}$ (\ref{7}) for a system with 6 protons and 6 neutrons distributed over 6 orbits.
Dashed lines show the energies of the lowest three pairs $\Gamma^\dag_i$ resulting from the diagonalization of $H^{(iv)}$ in the space spanned by the states $P^\dag_{kM_T}|0\rangle$ (see text). All values are in units of the level spacing $d$.}
\label{fig3}
\end{figure}

Due to the isospin invariance of $H^{(iv)}$, Eqs. (\ref{5}) and (\ref{6}) 
remain unaltered in the isovector case so that it is still true that $E_\nu \neq\lambda_\nu$,
in general. 
In Fig. \ref{fig3}, the solid lines represent the energies $E_\nu$ of the six pairs $B^\dag_\nu$ defining the ground state of the system. 
An important difference with respect to Fig. \ref{fig1} is that here two solid lines approach the same value $2\epsilon_i$ for $g\rightarrow 0$.  This is due to the fourfold degeneracy of the single-particle levels which allows two pairs to occupy the same orbit. Another difference is that the bottom two energies remain always distinct for this particular system \cite{sasa_jpg}. Apart from that, one may see important analogies between these two figures. Thus, in both figures
one can notice that the behavior of the energies $E_\nu$ follows closely  that of the  parameters $\lambda_\nu$ \cite{sasa_jpg}. In Fig. \ref{fig3} the
top four energies become equal two by two at the critical points where also the $\lambda_\nu$'s become equal by remaining real afterwards, a feature which is also seen in Fig. \ref{fig1}. Still in Fig. \ref{fig3}, the three dashed lines represent the energies of the lowest three collective pairs 
$\Gamma^\dag_\nu$ resulting from the diagonalization of $H^{(iv)}$ in the space spanned by the states $P^\dag_{kM_T}|0\rangle$. Only two of the three dashed lines are visible because the energy of the pair $\Gamma^\dag_1$ exactly overlaps $E_1$
(the solid black line) for the whole range of $g$ shown in the figure. An analogous overlap can be seen also for the energies of the pairs $\Gamma^\dag_2$ and $\Gamma^\dag_3$ but only within limited ranges of the strength. As in Fig. \ref{fig1}, there is a weak-coupling region, approximately $0\leq g\leq 0.2$, where corresponding solid and dashed lines overlap each other, what denotes a close identity between the pairs $B^\dag_\nu$ and the lowest pairs $\Gamma^\dag_\nu$ in this region. 
\begin{figure}
\includegraphics{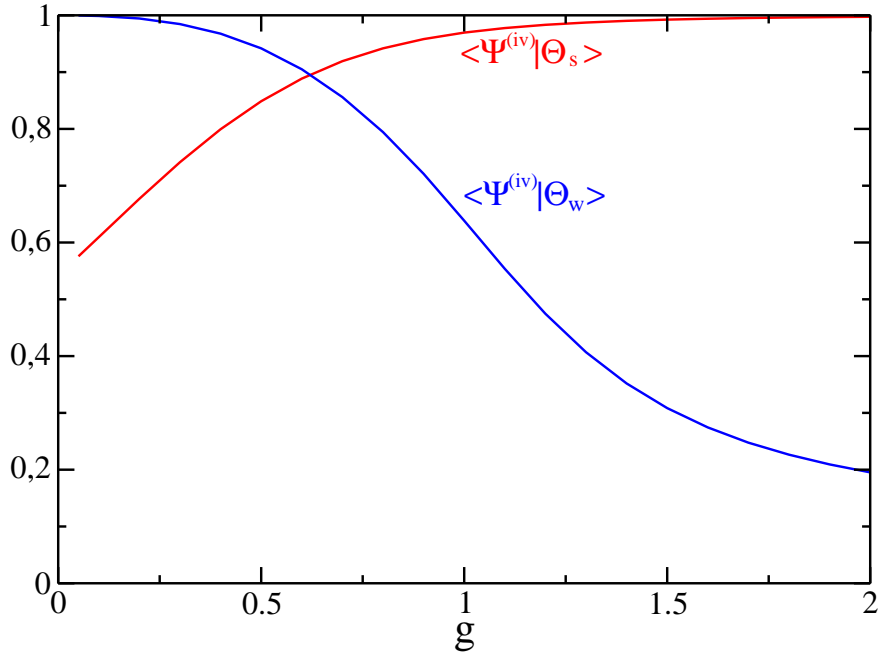}
\mbox{}\\[9.0cm]
\caption{The overlaps $\langle\Psi^{(iv)} |\Theta_w\rangle$ and $\langle\Psi^{(iv)} |\Theta_s\rangle$
discussed in the text as a function of the strength $g$ (in units of the level spacing $d$).}
\label{fig4}
\end{figure}

As for the like-particle pairing, the pairs $\Gamma^\dag_\nu$ can be used to build a new basis in which to expand the eigenstates of $H^{(iv)}$. In Fig. \ref{fig4}, we show the overlaps between the exact ground state $|\Psi^{(iv)}\rangle$ and two states (properly normalized) of this basis, namely
$|\Theta_w\rangle =[\Gamma^\dag_1\Gamma^\dag_1]^{0}[\Gamma^\dag_2\Gamma^\dag_2]^{0}
[\Gamma^\dag_3\Gamma^\dag_3]^{0}|0\rangle$ and
$|\Theta_s\rangle =([\Gamma^\dag_1\Gamma^\dag_1]^0)^3 |0 \rangle $, where the pairs $\Gamma^\dag_\nu$ have
been ordered for increasing values of their energy and the symbol $[\cdot\cdot\cdot]^0$ stands for coupling to $T=0$. 
It can be observed that the state $|\Theta_s\rangle$ has the structure of a quartet condensate, with the quartet
defined by $Q^+ = [\Gamma^\dag_1\Gamma^\dag_1]^0$.
As seen in Fig. \ref{fig4}, the overlap $\langle\Psi^{(iv)} |\Theta_w\rangle$ rapidly decreases from unity with increasing $g$ while 
the opposite is seen to happen to $\langle\Psi^{(iv)} |\Theta_s\rangle$. This testifies the evolution of the ground state towards a condensate of $T=0$ quartets formed by the lowest pair  in energy resulting from the diagonalization of $H^{(iv)}$ in the space spanned by the states $P^\dag_{kM_T}|0\rangle$. As for the like-particle pairing, this behavior 
reflects the rapidly increasing gap between $\Gamma^\dag_1$ and the remaining pairs.

\begin{figure}
\includegraphics{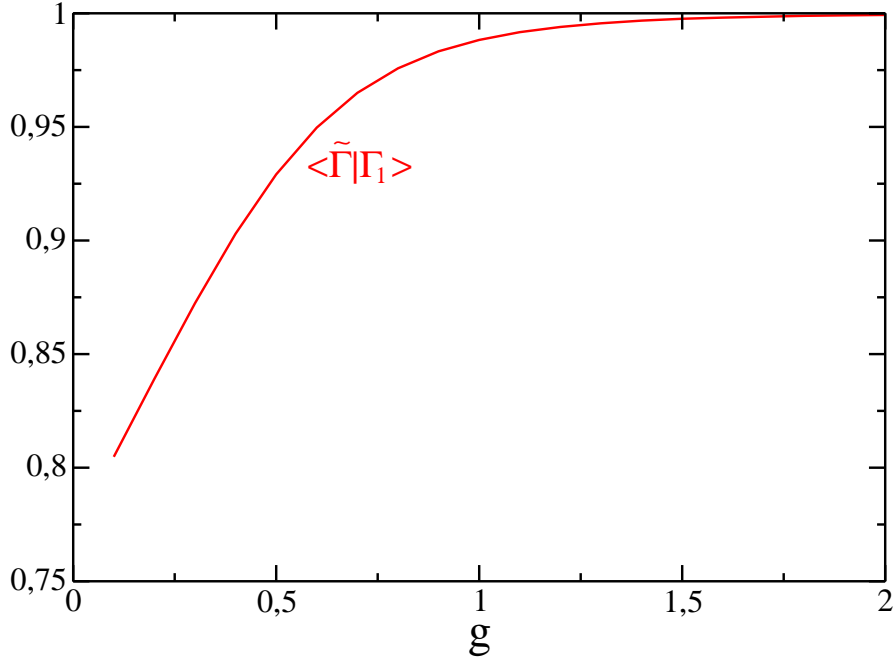}
\mbox{}\\[9.0cm]
\caption{The overlap between the QCM pair ${\tilde{\Gamma}}^\dag$ (\ref{14}) and the pair $\Gamma^\dag_1$
discussed in the text as a function of the strength $g$ (in units of the level spacing $d$).}
\label{fig5}
\end{figure}
\begin{figure}
\includegraphics{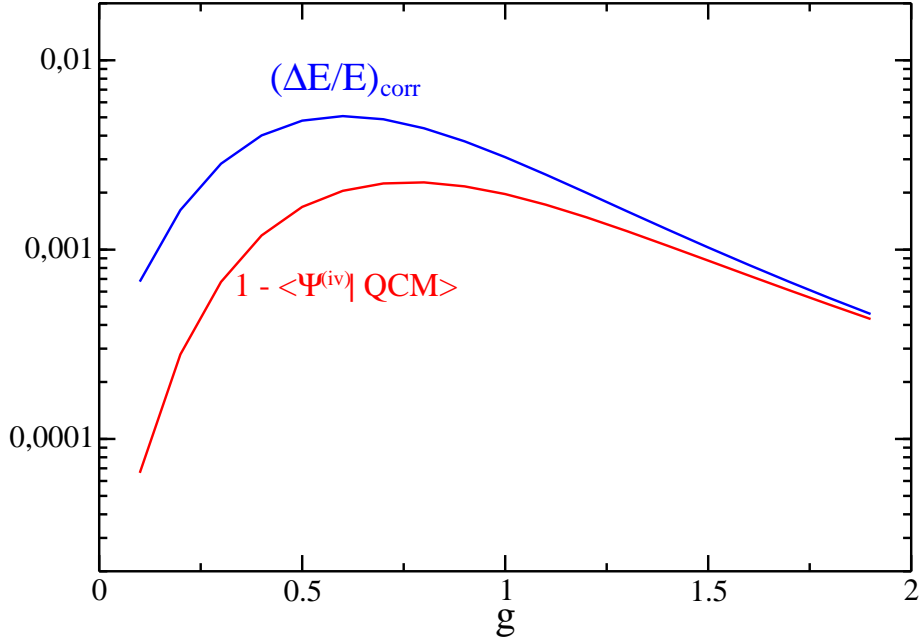}
\mbox{}\\[9.0cm]
\caption{Relative errors $(\Delta E/E )_{corr}$ in the QCM ground state correlation energy and deviations from unity of the overlap between exact and QCM ground states as a function of the strength $g$    (in units of the level spacing $d$).}
\label{fig6}
\end{figure}

It is worthy noticing that
the state $|\Theta_s\rangle$ does not represent the best approximation to the ground state that a condensate of this form can provide. The reason for that has to be searched in the way the pairs forming the quartets are constructed. These pairs arise from a diagonalization of the Hamiltonian in a space of two particles and therefore they are not affected   by the presence of the other pairs of the system. The accuracy of the condensate ansatz for the ground state can increase considerably by choosing these pairs such to guarantee the minimum energy  of the condensate, as required by the QCM \cite{sandu}. We recall that the QCM condensate is defined by
\begin{equation}
|QCM \rangle = (\tilde{Q}^\dag)^{n_q} |0 \rangle \equiv ([\tilde{\Gamma}^\dag \tilde{\Gamma}^\dag]^{T=0})^{n_q} |0 \rangle,
\label{14}
\end{equation}
where $\tilde{\Gamma}^\dag_{M_T}= \sum_i x_i P^\dag_{i M_T}$ are collective isovector pairs 
defined by the mixing amplitudes $x_i$. The latter are determined by minimising the average of the Hamiltonian (\ref{7}) in the trial state (\ref{14}). 
The QCM quartet $\tilde{Q}^\dag$ and the quartet $Q^\dag$ employed in the state $|\Theta_s\rangle$ become equivalent to the extent that the two pair operators $\Gamma^\dag_1$ and $\tilde{\Gamma}^\dag$ approach each other. As one sees in Fig. \ref{fig5}, which reports the overlap between these two pairs, this is indeed what happens with increasing $g$. 
The accuracy of the QCM wave function for the system studied here can be observed in Fig. \ref{fig6} where we show both the deviation from unity of its overlap with the exact ground state and the relative error in its correlation energy (the latter being  the difference between the actual energy of this state and its energy in the absence of the interaction). Both quantities are well below $1\%$ throughout the whole range of $g$ in the figure. 

As a final comment, we like to remark that, while remaining in a quartet formalism, the already high degree of accuracy of Fig. \ref{fig6} can be further increased by releasing two constraints present in the QCM wave function (14), namely letting the ground state to be a product of distinct quartets and, in addition, constructing each quartet not by two collective $T=1$ pairs but, 
more generally, as a linear superposition of two uncorrelated $T=1$ pairs. Applications of this approach can be found in Refs. \cite{sasa_prc,qm_t0t1,sasa_t01}.

\section{Conclusions}
In this work we have investigated the mechanism of quartet condensation in the ground state of 
even-even $N=Z$ systems governed by an isovector pairing force. For this purpose we have studied the evolution of this ground state from an unperturbed regime up to a strongly interacting one in a formalism of collective pairs. These pairs are those resulting from the diagonalization of this Hamiltonian in a space of two particles coupled to isospin $T=1$. The ground state has been seen to rapidly evolve from a product of distinct $T=0$ quartets, each formed by two of the above pairs, towards a condensate of quartets formed by the lowest pair in energy. Such an evolution closely follows the increasing gap in energy between this pair and the remaining ones. This gap, which reflects the unique properties of the lowest pair,  makes this quartet condensate to become predominant over all other states built with these pairs. The analogy with the mechanism responsible for the development of a pair condensate in the ground state of a like-particle pairing Hamiltonian has been underlined. This work establishes a link between the complicated structure of the exact ground state of 
the isovector pairing Hamiltonian and the simple approach of Quartet Condensation Model, which had remained unexplored so far.

\vskip 0.3cm
\begin{acknowledgments}
This work was supported by a grant of Romanian Ministry of Research and Innovation, CNCS - UEFISCDI, project number
PN-III-P4-ID-PCE-2016-0481, within PNCDI III.
\end{acknowledgments}


\end{document}